\documentclass[10pt,conference]{IEEEtran}
\usepackage[switch,columnwise]{lineno}

\IEEEoverridecommandlockouts
\usepackage{cite}
\usepackage{url}
\usepackage{amsmath,amssymb,amsfonts}
\usepackage[version=4]{mhchem}
\usepackage{chemformula}
\usepackage{chngcntr}
\newcommand{\R}{\mathbb{R}}
\newtheorem{definition}{Definition}
\newtheorem{Theorem}{Theorem}

\usepackage{algorithmic}
\usepackage{graphicx}
\usepackage{textcomp}
\usepackage{xcolor}
\def\BibTeX{{\rm B\kern-.05em{\sc i\kern-.025em b}\kern-.08em
    T\kern-.1667em\lower.7ex\hbox{E}\kern-.125emX}}
\begin{document}

\title{Formal characterization and efficient verification of a biological robustness property\\
}

\author{
\IEEEauthorblockN{Lucia Nasti}
\IEEEauthorblockA{\textit{Department of Computer Science} \\
\textit{University of Pisa}\\
Pisa, Italy \\
lucia.nasti@di.unipi.it}
\and
\IEEEauthorblockN{Roberta Gori}
\IEEEauthorblockA{\textit{Department of Computer Science} \\
\textit{University of Pisa}\\
Pisa, Italy \\
gori@di.unipi.it}
\and
\IEEEauthorblockN{Paolo Milazzo}
\IEEEauthorblockA{\textit{Department of Computer Science} \\
\textit{University of Pisa}\\
Pisa, Italy \\
milazzo@di.unipi.it}

}

\maketitle

\begin{abstract}

Robustness is an observable property for which a chemical reaction network (CRN) can maintain its functionalities despite the influence of different perturbations. In general, to verify whether a network is robust, it is necessary to consider all the possible parameter configurations. This is a process that can entail a massive computational effort. In the work of Rizk et al., the authors propose a definition of robustness in linear temporal logic (LTL) through which, on the basis of multiple numerical timed traces obtained by considering  different parameter configurations, they  verify the robustness of a reaction network. In this paper, we focus on a notion of initial concentration robustness ($\alpha$-robustness), that is related to the influence of the perturbation of the initial concentration of one species (i.e., the input) on the concentration of  another species (i.e., the output) at the steady state.  We characterize this notion of robustness in the framework proposed by Rizk et al., and we show that, for monotonic reaction networks, this allows us to drastically reduce the number of traces necessary to verify robustness of the CRN.

\end{abstract}


\section{Introduction}

Two main characteristics define living cells: an intrinsic structural complexity, according to which there is modularity inside the cell itself, and the ability to interact with other networks, working as a system. In addition to this already intricate framework, at various frequencies and timescales, internal and external fluctuations can alter specific functions or traits of biological systems, causing genetic mutations, loss of structural integrity, diseases and so on. Nevertheless, many biological networks can maintain their functionalities despite perturbations: this distinct property is known as \textit{robustness} \cite{kitano2004biological}.

Robust traits are pervasive in biology: they involve various structural levels, such as gene expression, protein folding, metabolic flux, species persistence. For this reason, the study of robustness is essential for biologists, whose aim is to understand the performance and functions of a biological system. 

However, it is not easy  to investigate biological systems since  they often  exhibit non-linear and non-intuitive behaviors. They can be studied by performing wet-lab (\textit{in vitro}) experiments, or through mathematical or computational (\textit{in silico}) methods on a pathway model. Unfortunately, the applicability of the last  approach is often hampered by the complexity of the models to be analyzed (often expressed in terms of ODEs or Markov chains). An alternative way is to infer properties of  the system by  analyzing  only  its structure, without studying or simulating its dynamics, as proposed in \cite{shinar2010structural}. However, this approach has been shown to be limited in its applicability because the derived conditions are often very restrictive.

Robustness can be formally studied by applying the methodology proposed by Rizk et al. in \cite{rizk2009general,rizk2011continuous}. Such a methodology is based on the definition of robustness given by Kitano in \cite{kitano2004biological} as {\em the ability of a system to maintain specific functionalities against perturbations}. The robustness of a system is measured as the {\em distance} between  the system behaviour under perturbations and its reference behaviour expressed as a linear temporal logic (LTL) formula. The distance is computed by using a notion of {\em violation degree} that measures  how much the temporal logic formula should be changed in order to match traces of perturbed behaviours obtained, for instance, through many simulations.

The approach proposed by Rizk et al. is very general, both in the description of the reference behaviour and in the  kind of perturbations considered. On the contrary, we want to focus on a  particular notion of robustness, namely the \textit{initial concentration robustness} that studies the influence of the initial concentrations of species on the concentration of  the
other species at steady state of the system. For these reasons,  we consider the notion of $\alpha$-robustness, based on continuous Petri nets \cite{gilbert2006petri} and interval markings, which extends the notion of {\em absolute concentration robustness} considered in \cite{shinar2010structural,shinar2011design}. 

The evolution  of a biological system under initial concentration perturbations can be understood by studying the  simulations of  the systems under all possible combinations of its initial concentrations, a process that generally entails a massive computational effort. In order to drastically reduce the number of simulations needed for this study, we proved that - if the concentration of an output species is monotonic with respect to the concentration of an input species (i.e. the perturbed element) – the number of simulations can be reduced to two by studying the model only on the extreme values of the input concentration range. This result  was the subject of the \textit{Input-Output theorem} on monotonicity~\cite{nasti2020mono}.
 
In this paper, we show that $\alpha$-robustness, proposed by Nasti et al.~\cite{nasti2018formalizing}, is a particular instance of the robustness notion proposed framework proposed by Rizk et al., and, therefore, it can be
characterized in their framework. The great advantage in case of a CRN  showing a monotonic behaviour according to our Input-Output theorem, this allows us to  drastically reduce the number of traces necessary to verify robustness.

The paper is organized as follows.
We proceed by first introducing the notion of Chemical Reaction Networks (CRN) in Section \ref{CRN}. In Section \ref{LTL}, we give the formal definition and the semantics of Linear Temporal Logic, which represents the base of the work done by Rizk et al, described in Section \ref{fages}. In Section \ref{our_def}, we give the formal definitions of initial concentration robustness  together with the characterization of $\alpha$-robustness in terms of the framework proposed by Rizk et al.. We also characterize monotonicity in CRN. 
We apply our approach to the biological example described in Section \ref{example}. Finally, in  Section~\ref{conclusion} we draw some conclusions.

    
\section{Background}
 We introduce some notions that will be assumed in the rest of the paper. In the first part, we focus on the representation of chemical reactions, considering one of the main methods that we can use to describe them: the \textit{deterministic approach}. In the second part, we present the Linear Temporal Logic formalism that provides a mathematical notion to express systems behavior, and it is at the basis of the work done by Rizk et al.. 

\subsection{Chemical Reaction Networks} \label{CRN}
\noindent A chemical reaction is a transformation that involves one or more chemical species, in a specific situation of volume and temperature.

We call \textit{reactants} the chemical species that are transformed; while those that are the result of the transformation are called \textit{products}. We can represent a chemical reaction as an equation, showing all the species involved in the process. 

A simple example of chemical reaction is the following elementary reaction:
\begin{equation}\label{eq:two_arrows}
\ch{ \textit{a}A + \textit{b}B <=>[$k\sb{1}$][$k\sb{-1}$]\textit{c}C + \textit{d}D }
\end{equation}
In this case, A, B, C, D are the species involved in the process: A and B are the reactants, C and D are the products. The parameters \textit{a}, \textit{b}, \textit{c}, \textit{d} are called \textit{stoichiometric coefficients} and represent the number of reactants and products participating in the reaction. The arrow is used to indicate the direction in which a chemical reaction takes place. When we have only one arrow, it means that the reaction is \textit{irreversible}, that is it is not possible to have the opposite process. To describe the dynamical behaviour of the chemical reaction network, we can use the \textit{law of mass action}, which states that: the rate of a reaction is proportional to the product of the reactants. Applying the \textit{law of mass action} to the system, we obtain, for each chemical species, a differential equation describing the production and the consumption of the considered species. Considering the generic chemical equation \ref{eq:two_arrows}, we obtain:
\begin{align*}
\begin{split}
&\frac{d[A]}{dt}=\overbrace{-a k_{1} [A]^{a}[B]^{b} \,}^{\parbox{5em}{\centering\tiny{direct reaction\\ term}}} \, \overbrace{ +a k_{-1} [C]^{c} [D]^{d}}^{\parbox{5em}{\centering\tiny{inverse reaction\\ term}}}\\
&\frac{d[B]}{dt}=-b k_{1} [A]^{a}[B]^{b} + b k_{-1} [C]^{c} [D]^{d}\\
&\frac{d[C]}{dt}=+c k_{1} [A]^{a}[B]^{b} - c k_{-1} [C]^{c} [D]^{d}\\
&\frac{d[D]}{dt}=+d k_{1} [A]^{a}[B]^{b} - d k_{-1} [C]^{c} [D]^{d}.
\end{split}
\end{align*}
where, in each equation, we isolated the term describing the direct reaction from the one describing the inverse reaction. With these two terms, we implicitly considered, for each element, the processes of consumption and production.



\subsection{Formal definition of LTL} \label{LTL}

Linear Temporal Logic is a logical formalism. It provides a mathematical notion to express systems behaviors \cite{baier2008principles}, based on a linear-time perspective. The temporal logic is necessary to specify the relative order of events, expressed by elementary modalities, which combined can express complex dynamical properties. Typical properties are oscillations (when a behavior recurs infinitely), reachability (when the system can reach a given state), invariance (when a property is always true), inevitability (when a system has to reach a given state), response (an event causes a specific behavior) \cite{emerson1990temporal}.

A basic LTL formula $\phi$ consists of atomic propositions $a \in AP$, Boolean connectors ($\land$, $\lor$, $\lnot$, $\implies$), and two basic modal operators:
\begin{itemize}
    \item $\mathbb{X}\phi$ or $\circ$ (``next") means that a given formula $\phi$ is true in the next state;
    \item $\phi\mathbb{U}\phi$ or $\cup$ (``until") means that given two formulas $\phi_1$ and $\phi_2$, the formula $\phi_1$ is true, until the formula $\phi_2$ becomes true.
\end{itemize}
The ``until" operator $\cup$ allows to derive other two temporal modalities, defined as follows:
\begin{itemize}
	 \item $\mathbb{F}\phi$ or $\diamond$ ``eventually (in the future)" means that a given formula $\phi$ is true now or sometime in the future, defined as $\diamond \phi \overset{\mathrm{def}}{=} true \cup \phi$;
    \item $\mathbb{G}\phi$ or $\square$ ``globally" means that a given formula $\phi$ is true now and forever, defined as $\square \phi \overset{\mathrm{def}}{=} \neg \diamond \neg \phi$.
\end{itemize}

The LTL formulae are formed according to the following grammar:  

\begin{center}
$\phi:: $ $true |$ $ a |$ $ \phi_1 \land \phi_2 |$  $ \lnot \phi |$ $\circ \phi |$ $\phi_1 \cup \phi_2|$ 
\end{center}
The atomic proposition $a$, with $a \in AP$ is a state label, representing an assertion about the value of a system variable that has to be evaluated, such as the concentration of a chemical species. 

\begin{figure}[h]
\centering
  \includegraphics[width=0.85\linewidth]{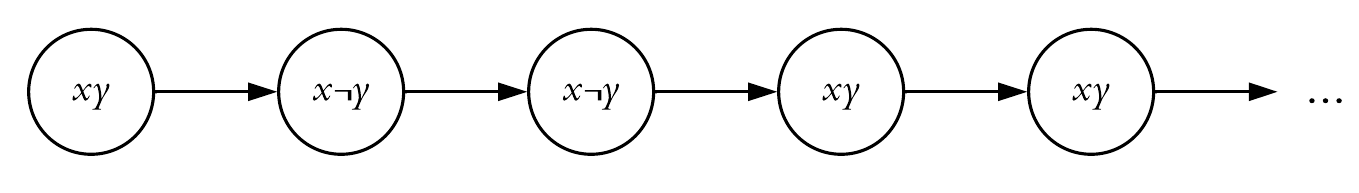}
\caption{Example of Linear Temporal Logic.}
\label{fig: ltl_example}
\end{figure}

A LTL formula $\phi$ represents a property of a trace, which is a infinite path. Given a path and a formula $\phi$, we can formulate precisely when $\phi$ holds on the path. For example, the trace $T$ in Figure \ref{fig: ltl_example} satisfies the formula $\phi_1= \mathbb{F}(x \lor \lnot y)$ because it is true in the first state of the trace $T$. Instead, the trace does not satisfy the formula $\phi_2= \mathbb{G} (y)$ because $y$ is not true in all states.

\subsubsection{Semantics of LTL}
\begin{figure}[h]
\centering
  \includegraphics[width=0.99\linewidth]{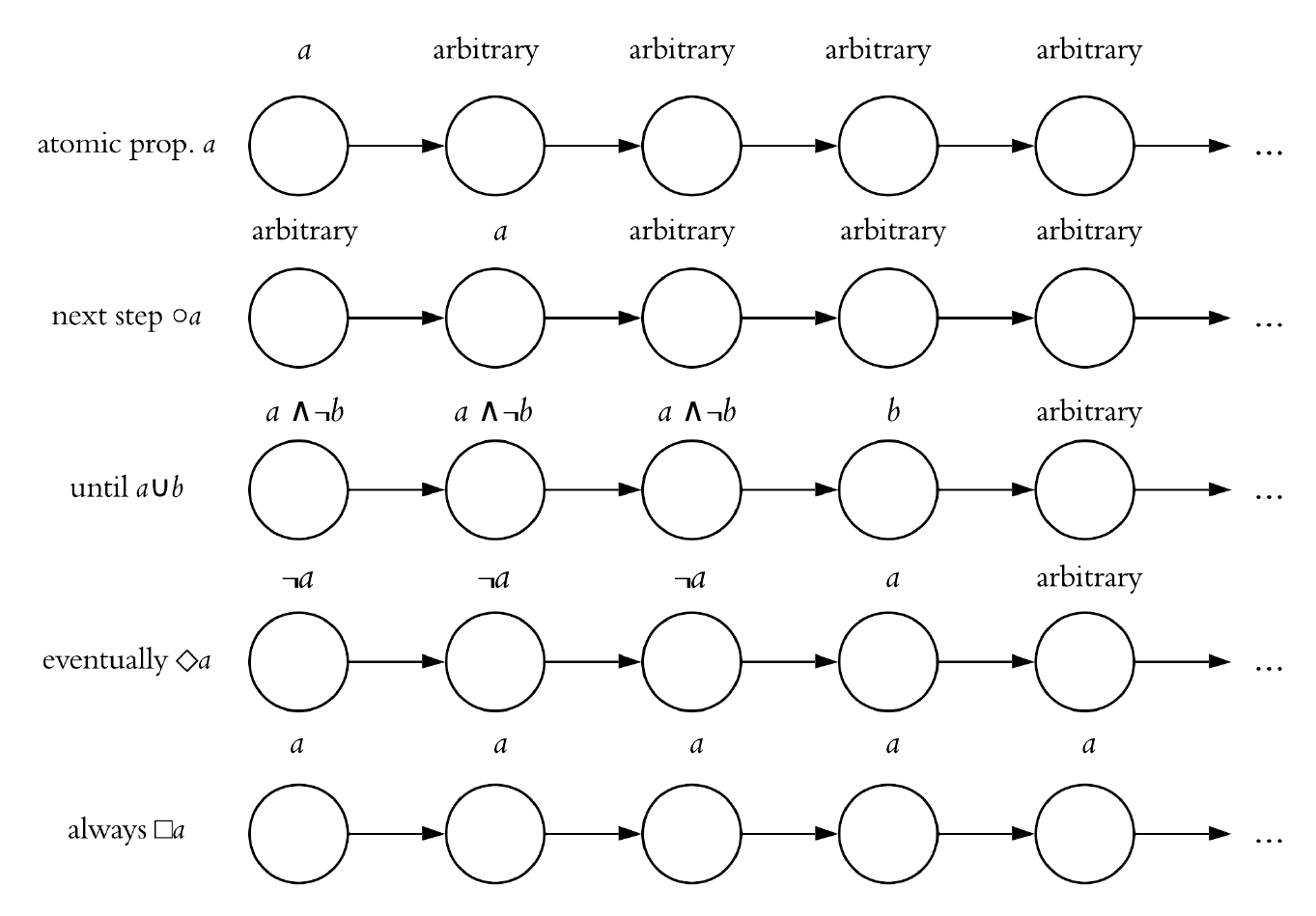}
\caption{Intuitive sketch of LTL semantics.}
\label{fig: ltl_sem}
\end{figure}

To precisely formulate when a path satisfies an LTL formula, we define the semantics of LTL formula $\phi$ by providing a satisfaction relation $\models$ such that $\sigma \models \phi$ if and only if a property $\phi$ is satisfied by a trace $\sigma$. \cite{baier2008principles}:

\begin{definition}
Let $\phi$ be an LTL formula over $AP$ and $\sigma \in (2^{AP})^{\omega} $ be a trace. The satisfaction relation $\models$ $\subseteq$ $(2^{AP})^{\omega} \times LTL $ is the smallest relation with the following properties: 

\begin{align*}
\begin{split}
&\sigma \models true\\
&\sigma \models a \quad iff \quad a \in A_0  \quad (i.e., A_0 \models a)\\
&\sigma \models \phi_1 \land \phi_2  \quad iff \quad \sigma \models \phi_1 \quad and \quad \sigma \models \phi_2\\
&\sigma \models \neg \phi \quad iff \quad \sigma \nvDash \phi\\
&\sigma \models \circ \phi \quad iff \quad \sigma[1...] = A_1A_2A_3... \models \phi\\
&\sigma \models \phi_1 \cup \phi_2 \quad iff \quad \exists j \geq 0.\sigma[j...] \models \phi_2 \quad and \quad\\ 
&\sigma[i...] \models \phi_1, \quad \forall 0\leq i < j.
\end{split}
\end{align*}

Here, for $\sigma = A_0 A_1 A_2...\in (2^{AP})^{\omega}$, $\sigma[j...] = A_j A_{j+1} A_{j+2}...$ is the suffix of $\sigma$ starting
in the $(j+1)$st symbol $A_j$.

For the derived operator $\diamond$ and $\square$ the expected result is: 
\begin{align*}
\begin{split}
&\sigma \models \diamond \phi \quad iff \quad \exists j \geq 0. \sigma[j...] \models \phi\\
&\sigma \models \square \phi \quad iff \quad \forall j \geq 0. \sigma  [j...] \models \phi.
\end{split}
\end{align*}
\end{definition}
In Figure \ref{fig: ltl_sem}, we add an intuitively sketch of the semantics of temporal modalities.

\section{Application of LTL to study Robustness} \label{fages}

\subsection{Temporal logic semantics of numerical traces} 
Numerical simulations are used to obtain the behavior of a biological system, which is described by a \textit{numerical timed trace} as shown in \cite{rizk2009general}. A \textit{numerical timed trace}, expressing the evolution of a system with time, is a finite sequence of tuples $T=(s_0, s_1,..., s_n)$ with $s_i = (t_i, \mathbf{x_i},\mathbf{\dot{x}_i})$ where $t_i$ with 
$i\in [0, n]$ is a sequence of increasing time points, $\mathbf{x_i}$ is the vector of state variable values and $\mathbf{\dot{x}_i}$ is the derivative of state variable at time $t_i$. 

\begin{figure}[!ht]
\centering
  \includegraphics[width=0.9\linewidth]{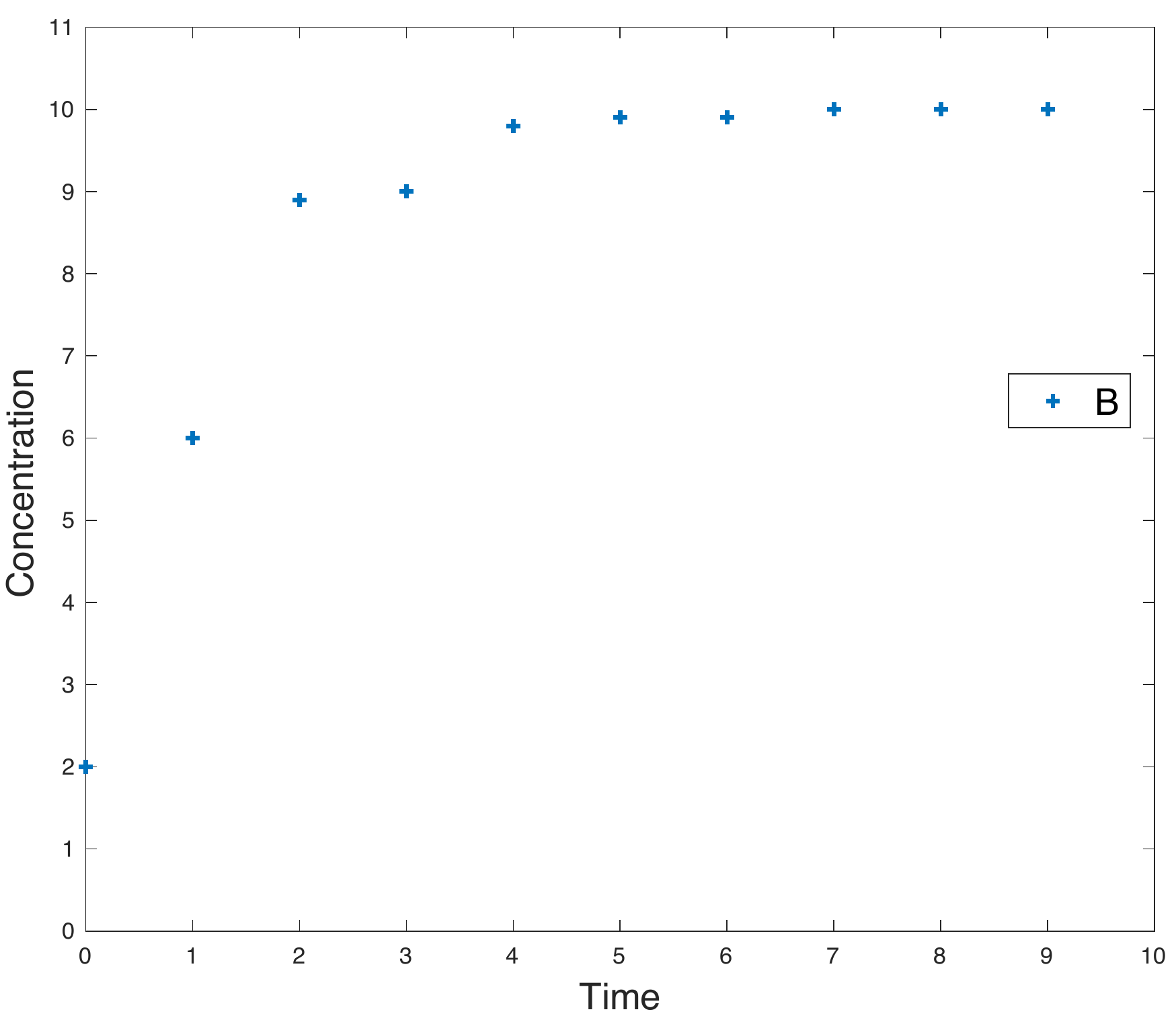}
\caption{Numerical trace representing the time evolution of the concentration of chemical species $B$. }
\label{fig: es_ch3}
\end{figure}

By a numerical trace, we can depict different biological phenomena, such as the time evolution of a concentration of a chemical species in a system, as represented in Figure \ref{fig: es_ch3}. In this example, $T=((0,2,0),(1,6,4.12),...,(9,10,0))$ is the the associated trace, in which each state variable defines a specific concentration level of the species $B$ over time.

In \cite{rizk2009general}, the authors use LTL to express dynamical properties of biological systems. LTL operators describe if and when a property $\phi$ holds on a trace $T$. Considering again Figure \ref{fig: es_ch3}, the formula $\phi = \mathbb{F}([B]>7)$ expresses that at some point the concentration of species $B$ is greater than $7$. 

\subsection{The formal definition of Robustness degree}
Since it is interesting to define how much a numerical trace satisfies a formula $\phi$, the authors specify the quantifier-free LTL (QFLTL), which replaces the numerical constants in the atomic propositions of a LTL formula, with free real-valued variables $\mathbf{y}$. Then, in this way, having a formula $\phi$ and a vector of real-valued variables $\mathbf{y}$, it is possible to know for which values $\mathbf{y}$ the QFLTL formula $\phi(\mathbf{y})$ holds on $T$. At this point, the \textit{satisfaction domain} is defined as follows:

\begin{definition}[Satisfaction Domain]	\label{domain}
Given a QFLTL $\phi$ formula, for any trace $T$, the satisfaction domain of $\phi(\mathbf{y})$ is the set of variables $\mathbf{y}$ for which $\phi(\mathbf{y})$ holds. It is defined as:  

\begin{equation}
D_{T,\phi(\mathbf{y})} =  \{ \mathbf{y} \in \mathbb{R}^q | T  \models \phi (\mathbf{y}), \}
\end{equation}
where $q$ is the number of constants appearing in $\phi$. 
\end{definition} 

Through this approach, the LTL formula becomes an instance of a more general QFLTL formula obtained by variable abstraction. Considering again Example \ref{fig: es_ch3}, and the formula $\phi = \mathbb{F}([B]>2 \land \mathbb{F}[B] < 10)$, it is possible to associate the formula $\phi(\mathbf{y}) = \phi(y_1, y_2) = \mathbb{F}([B]>y_1 \land \mathbb{F}([B]< y_2))$. Moreover, concerning trace $T$ in Example \ref{fig: es_ch3}, the domain is $D_{T,\phi(y_1, y_2)}=\{y_1 \leq 10 \land y_2 \geq 2 \} $, since $2$ and $10$ are respectively the minimum and the maximum values of the trace. 

Given a trace $T=(s_0, s_1,...,s_n)$ and a LTL formula $\phi$, the authors define the notion of \textit{violation degree} to quantify how much $\phi$ must be changed to hold on $T$. This concept is defined as the \textit{Euclidean distance} between a formula $\phi$ and the domain of the trace $T$. It is formally defined as follows:

\begin{definition}[Violation Degree]\label{vd}
The violation degree $vd(T, \phi)$ of a formula $\phi$ with respect to a trace $T$ is the distance between the actual specification and validity domain $D_{T, \phi(\mathbf(y))}$ of the QFLTL formula $\phi(\mathbf{y})$ obtained by variable abstraction:
\begin{equation}
vd(T, \phi) = dist(\phi, D_{T, \phi(\mathbf{y})})
\end{equation}
\end{definition}

Considering Example \ref{fig: es_ch3} and the formula $\phi_1 = \mathbb{F}([B]>2 \land \mathbb{F}([B]<10))$, by violation degree, we can compute how much $\phi$ is distant from the domain: in this case, we obtain $vd(T, \phi_1) = 0$ because $\phi$ is satisfied by $T$. Instead, if we consider the formula $\phi_2 =\mathbb{F} ([B]>12 \land \mathbb{F}([B]< 3) )$, the violation degree is $vd(T, \phi_2) = 2$ meaning that $\phi_2$ has to be changed to hold on $T$.

To define, instead, how much the given LTL formula holds on a given numerical trace, the notion of \textit{satisfaction degree} is introduced as follows: 

\begin{definition}[Satisfaction Degree]\label{sd}
 The satisfaction degree $sd(T, \phi)$ of a formula $\phi$ with respect to a trace $T$ is defined as: 
 \begin{equation}
 sd(T, \phi)= \frac{1}{1+vd(T, \phi)} \in [0,1],
 \end{equation}
 where $vd(T, \phi)$ represents the violation degree. 
\end{definition}

The value obtained by computation of the satisfaction degree ranges between $0$ and $1$. The satisfaction degree is equal to $1$ when the trace $T$ satisfies a formula $\phi$, otherwise it tends to $0$. The concept of satisfaction degree characterizes the definition of robustness.  

\begin{definition}[Rizk et al. Robustness] \label{def_robustness_Fages}
The robustness of a system is defined as: 
\begin{equation}
R_{\phi,P}^s= \int_{p \in P} prob(p) sd(T_p, \phi)dp
\end{equation}
where $\phi$ is the specification of the functionality in LTL; $T_p$ is the numerical trace, representing the system behavior under perturbation $p$; $P$ is the set of perturbations. 
\end{definition}

The continuous probability distribution characterizes the perturbations, affecting the entire system: each perturbation has its weight, representing how much it can influence the biological behavior under study. 

This entire approach is already implemented in BIOCHAM (BIOCHemical Abstract Machine)~\cite{calzone2006biocham}, a software environment for modeling biochemical systems, where the function computing robustness is the following: 
\begin{center}
 \texttt{robustness(LTL Formula, [parameters name], [variable objective])}.   
\end{center}

As described in~\cite{calzone2006biocham}, the function computes the robustness degree with respect to the \texttt{LTL Formula}, for the list of \texttt{[parameters name]} and with list of objectives for the free variables of \texttt{LTL Formula} given in \texttt{[variable objective]}.

The following two functions can be used respectively to set the coefficient of variation of the parameters and the number of simulations:

\begin{itemize}
    \item \texttt{option(robustness\_coeff\_var: number)};
    \item \texttt{option(robustness\_samples: integer)}.
\end{itemize}

\subsection{Example: Application to the ERK signaling pathway}\label{example}

As example, we show how to analyze the dynamics of the ERK signalling pathway.

A signalling pathway consists of enzymatic cascades, having a starting species that triggers the other connected reactions. In general, there is a particular species, namely the \textit{transductor} that perceives an initial stimulus, which activates the cascade amplifying the signal for the next enzymatic process. 

One of the most important examples of such processes is the ERK pathway, involved in many biological phenomena such as cell's growth and differentiation. We consider a particular portion of the mathematical model of the ERK pathway (denoted as ERK$*$) implemented by Schilling et al. \cite{schilling2009theoretical} and available to the public on the BioModels Database (BIOMD0000000270). We indicate the species and the kinetics rates as originally denoted in the model in \cite{schilling2009theoretical}. For simplicity, we refer to the reaction using the notation $R_i$, where $i$ is the kinetics rate index. The reactions involved are the following: 

\begin{align}
\begin{split}
&\ch{Raf <=>[$k\sb{18}$][$k\sb{19}$] PRaf} \\
&\ch{Mek1<=>[$k\sb{21}$[PRaf]][$k\sb{27}$]PMek1} \\
&\ch{PMek1<=>[$k\sb{23}$][$k\sb{25}$]PPMek1}, \\
\label{erk_reaction1}
\end{split}
\end{align}
in Table \ref{Table_example2} we reported the coefficient rates and the initial conditions of ERK$*$ system.

The species \ch{PRaf}, involved in the reaction $R_{21}$, acts as \textit{catalyst promoter}, which means that its concentration positively influences the production of the species PMek1. A catalytic species increases the reaction rate of the reaction in which it is involved, and during this process, its concentration is not consumed.

\begin{table}[t]
\caption{The initial concentrations and the rates of ERK$*$ system. }
\centering
\begin{tabular}{|l|l|}
\hline
Initial concentrations & Rates \\ \hline
Raf = $10$             & $k_{18}= 0.1445$      \\
Praf = $0$             &  $k_{19}= 0.37$      \\
Mek1= $1$              &  $k_{21}= 0.02$      \\
PMek1 = $0$            &  $k_{27}= 0.07$      \\
PPMek1 = $0$           &  $k_{23}= 667.957$      \\ 
                       &  $k_{25}= 0.13$          \\ \hline
\end{tabular}
\label{Table_example2}
\end{table}

\begin{figure}[h]
\centering
  \includegraphics[width=0.9\linewidth]{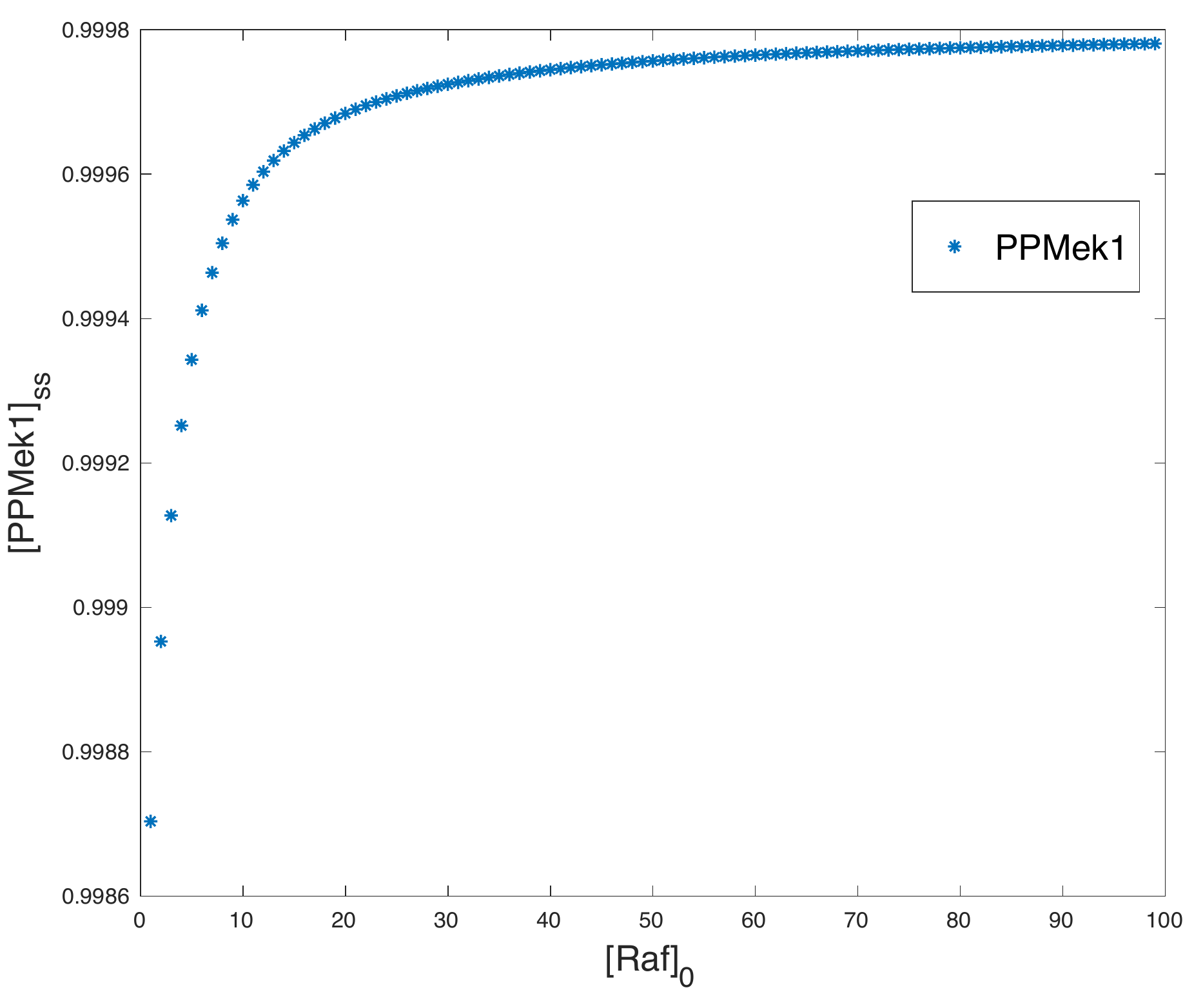}
\caption{Simulation results of CRN \ref{erk_reaction1}, representing ERK signalling pathway. To show how the concentration of the species PPMek1 varies at the steady state with respect to the initial concentration of the species Raf. We plot on the horizontal axis the initial concentration of Raf, in a range $[1, 100]$ and on the vertical axis the concentration of PPMek1 at the steady state.}
      \label{Figure_6}
      \end{figure}

To verify the robustness of the system, we need to perform many simulations, considering different kinds of perturbations. For instance, if we want to test whether the species PPMek1 is robust with respect to oscillations on the initial concentration of Raf, we need to compute many simulations, one for each possible (continuous) value of the initial concentration of Raf. In Figure~\ref{Figure_6}, we show how the concentration of the species PPMek1 varies at the steady state with respect to the initial concentration of the species Raf. We notice that PPMek1 does not exhibit \textit{absolute robustness}.
 
\section{The definition of the $\alpha$-robustness and its characterization in the Rizk framework} \label{our_def} 

The definition of robustness degree given by Rizk et al. represents the average satisfaction degree of the property of interest over all admissible perturbations, possibly weighted by their probabilities~\cite{rizk2011continuous}.

In our case, as already described by Nasti et al. in \cite{nasti2018formalizing}, we want to focus on the evaluation of the initial concentration robustness. We want to vary the concentration of at least one chemical species (namely the \textit{input}) and verifying, at the equilibrium, if the concentration of another species (namely the \textit{output}) is included in an interval of possible values. In order to do that, we recall the definition of the initial concentration robustness, using continuous Petri nets.

\begin{definition}[Continuous Petri net] A {\em continuous Petri net} $N$ can be defined as a quintuple
$\langle P , T, F, W, m_{0} \rangle$
where:
\begin{itemize}
    \item $P$ is the set of continuous \textit{places}, conceptually one for each considered kind of system resource;
    \item $T$ is the set of continuous \textit{transitions} that consume and produce resources;
    \item $F\subseteq (P \times T) \bigcup (T \times P) \rightarrow \mathbb R_{\ge 0}$ represents the set of arcs in terms of a function giving the weight of the arc as result: a weight equal to 0 means that the arc is not present;
    \item $W:F \rightarrow \mathbb{R}_{\ge 0}$ is a function, which associates each transition with a {\em  rate};
    \item $m_0$ is the {\em initial marking}, that is the initial  distribution of \textit{tokens} (representing resource instances) among places. A marking is defined formally as $m:P \rightarrow \mathbb R_{\ge 0} $.
\end{itemize}
\end{definition}
Tokens are movable objects, assigned to places, that are consumed by transitions in the input places and produced in the output places. Graphically, a Petri net is drawn as a graph with nodes representing places and transitions. Circles are used for places and rectangles for transitions. Tokens are drawn as black dots inside places. Graph edges represent arcs and are labeled with their weights. To  faithfully model biochemical networks, the marking of a place is not an integer (the number of tokens) but a positive real number (called {\em token value} representing the concentration of a chemical species. Each transition is associated with a kinetic constant, that determines the rate of (continuous) flow of tokens from the input to the output places of the transition.

In order to give the definition, we recall some definitions introduced by Nasti et al. in~\cite{nasti2018formalizing}. The initial marking is defined as an assignment of a fixed value to each place $p$. Now, it is possible to generalize the idea of initial marking by considering a marking as an assignment of a {\em interval of values} to each place $p$ of the Petri net. 

We first recall the definition of the domain of intervals.
\begin{definition}[Intervals]
The interval domain is defined as \[{ \cal I} = \{[n, m] \mid n, m \in \mathbb{R}_{\ge 0} \cup \{+\infty\} \mbox{ and } n\leq m \}.\]  An interval $[n,m] \in { \cal I}$ is {\em trivial} iff $n=m$. Moreover, $x \in [n, m]$ \text{iff} $n\leq x \leq m$.
\end{definition}
We now define interval markings.
\begin{definition}[Interval marking]
Given a set of places $P$, an {\em interval marking} is a function $m_{	[ \ ] \  } : P \rightarrow { \cal I}$. 
 The domain of all interval markings is $M_{	[ \ ] }$.
\end{definition}

An interval marking in which at least one interval is non-trivial represents an infinite set of markings, one for each possible combination of values of the non-trivial intervals. Therefore, given an interval marking, we relate it with the markings as in the original Petri nets formalism in the following way:
\begin{center}
Given $m \in M$ and $m_{[ \ ]} \in M_{[ \ ]}$, $m \in m_{[ \ ]}$ iff $\forall p \in P, m(p) \in m_{[ \ ]}(p) $.
\end{center}

In a Petri net we assume that there exists {\em at least one} input place and {\em exactly one} output place representing input and output species of the modeled biochemical network, respectively. Under this assumption, we can give the formal definition of robustness.

\begin{definition}[$\alpha$-Robustness]\label{defRobustness}
A Petri net $N$ with output place $O$ is \emph{$\alpha$-robust with respect to a given interval marking $m_{	[ \ ]}$} iff $ \exists  k \in \mathbb{R}$ such that $\forall m \in m_{	[ \ ]} $, the marking $m'$ corresponding to the steady state reachable from $m$, is such that
$$m'(O) \in [k- \frac{\alpha}{2}, k+ \frac{\alpha}{2} ]\ .$$
\end{definition}

\subsection{The initial concentration robustness in the general Rizk's framework} 

The specific notion of the initial concentration robustness is analyzable in the general framework proposed by Rizk et al.. To match the two frameworks, we restrict the analysis of the system behavior at the equilibrium. We show that Definition \ref{defRobustness} is an instance of Definition \ref{def_robustness_Fages}, expressed using the Rizk et al. framework shown in \cite{rizk2009general}.

\begin{Theorem}[Definition \ref{defRobustness} in the context of Definition \ref{def_robustness_Fages}]\label{equivalenza} 
Given a Petri Net $PN$ with output place O, an initial marking $m_{  [ \ ] \  } \in M_{  [ \ ] \ }$ and a continuous probability distribution $prob(m)$, defined on  $m_{  [ \ ] \  }$ such that the integral of the pdf is normalized to 1. $PN$ is $\alpha$-robust with respect to $m_{  [ \ ] \  }$ iff there exists an interval $[min, max] \in \R$  such that $R^{s}_{\phi,P}=1$, with $\phi = F(G([O]\geq min \land [O]\leq max ))$, $P$ equivalent to $m_{  [ \ ] \  }$ and $max-min=\alpha$.
\end{Theorem}

Hence, according to Theorem \ref{equivalenza}, we define the initial concentration robustness as: 

\begin{equation} \label{our_definition}
R_{\phi,m_{  [ \ ] \  }}^s= \int_{m \in m_{  [ \ ] \  }} prob(m) sd(T_m, \phi)dm.
\end{equation}
%


We can implement the Formula~\ref{our_definition} in BIOCHAM:

\begin{center}
    \texttt{robustness(F(G([O] >= min $\land$ [O] <= max)), [In], [min -> x, max -> y]),}
\end{center}

where:

  \begin{itemize}
    \item \texttt{robustness} is the function, implemented in BIOCHAM \cite{calzone2006biocham}, computing the robustness measure;
    \item \texttt{In} is the concentration of the input species, which is perturbed;  
    \item \texttt{O} is the concentration of the output species; 
    \item \texttt{F} and \texttt{G} are two specific temporal operators, respectively meaning \textit{future} and \textit{globally};
    \item the expression \texttt{[min -> x, max -> y]} represents the assignment of the interval limits;
    \item the expression \texttt{F(G([O] >= min $\land$ [O] <= max))} represents the system behavior that has to be evaluated. The formula expresses that, when the system reaches the steady state, the concentration of output species is within a range.
  \end{itemize}

The definition of $\alpha$-robustness is simpler and much less general than the one considered by Rizk et al.. However, it is conceived with the aim of enabling further studies on sufficient conditions that could allow robustness to be assesses by avoiding (or significantly reducing) the number of simulations to be performed. This could be obtained, for instance, by adapting conditions already considered in the context of monotonicity analysis \cite{angeli2006structural}, as we will describe in the next Section. 

\subsection{Input-Output Monotonicity in CRN}
In general, the verification of robustness requires a huge number of simulations~\cite{shinar2010structural}, being necessary to test the system behavior for all the possible combinations of initial concentrations of the involved chemical species.

In order to reduce the computational effort, we apply a sufficient condition that allows us to analyze the monotonicity between the \textit{output} of the system (the under study chemical species at the steady state) with respect to the initial concentration of the \textit{input} (the perturbed chemical species)~\cite{nasti2020mono}. In this context, the input and output are in a monotonicity relation if the concentration of output at any time either increases or decreases due to an increase in the initial concentration of the input, within an interval $[min, max] \in \R$:

\begin{definition}[Positive (respectively Negative) Input-Output Monotonicity]
Given a set of reactions $\mathcal{R}$, species $\mathcal{S}_O$ is \emph{positively monotonic} (resp. \emph{negatively monotonic}) with respect to $\mathcal{S}_I$ in $\mathcal{R}$ if, for any two initial states $S^0, \overline{S^0}$ as above, $\overline{S_O}(t) \geq S_O(t)$ (resp. $\overline{S_O}(t) \leq S_O(t)$), for every time $t\in \R_{\geq 0}$.
\end{definition}

If the monotonicity is established, we can consider only the extreme values of the input concentration interval, as described in~\cite{gori2019towards} by Gori et al.. Indeed, as we will see in detail in Section~\ref{example2}, if the output is monotonic with respect to the input, then just two simulations are necessary: one with input $=min$ and one with input $=max$. The dynamics of the output in all the other (intermediate) cases is included in the results we obtained from these two configurations.

The proposed condition is based on a constraint on the structure of the chemical reaction network that can be evaluated efficiently, without the need of performing simulations. Following the guidelines of~\cite{angeli2006structural}, it is based on a graph representation of the chemical reaction network, namely the \textit{R-graph}, enriched with information about cooperation and competition among reactions (i.e., the \textit{Labelled R-graph}). 

The R-graph is formally defined as follows.
\begin{definition}[R-graph]\label{l_g_def}
Given a finite set of reactions $\mathcal{R}$ over a set of species $\mathcal{S}$, the R-graph of $\mathcal{R}$ is the signed graph $\langle \mathcal{R}, E_{+}, E_{-} \rangle$, where $E_{+} \subseteq (\mathcal{R} \times \mathcal{R}) $ and  $E_{-} \subseteq (\mathcal{R} \times \mathcal{R})$ are defined as follows:
\begin{itemize}
    \item $(\mathcal{R}_i, \mathcal{R}_j) \in E_{+} $ if $i\neq j$ and there is a species which is product of $\mathcal{R}_i $ and reactant in $\mathcal{R}_j$;
     \item $(\mathcal{R}_i, \mathcal{R}_j) \in E_{-} $ if $i\neq j$ and there is a species which is a reactant (or product) in both reactions $\mathcal{R}_i$ and $\mathcal{R}_j$.
\end{itemize}
\end{definition}

Intuitively, given two reactions, we draw a positive edge between them if they \textit{cooperate} each other, hence, for example, the product of a reaction is among the reactants of the other reaction. Instead, we draw a negative edge between two reactions if both share the same reactants, hence they \textit{compete}.

In the R-graph, essentially, each edge is associated with one of the symbols $+$ or~$-$. We can label in the same way also its vertices, assigning to each one a sign $+$ or~$-$:

\begin{definition}[Consistent labeling]
A map $\sigma: \mathcal{R} \to \{+,-\}$ is called a \emph{consistent labeling} of the R-graph if the following two properties hold:
\begin{itemize}
    \item For each $(i,j) \in E_+$,  $\sigma(i)$ is equal to $\sigma(j)$.
    \item For each $(i,j) \in E_-$, $\sigma(i)$ is the opposite of $\sigma(j)$.
\end{itemize}
\end{definition}

The condition is then expressed as a set of constraints on the graph structure. In addition, studying the signs of the stoichiometric matrix of the network, it is possible to predict if a variation on the concentration of the input affects the output positively or not, as described in~\cite{nasti2020mono}, where we find the main result: 
\begin{Theorem} \label{THEOREM}
Given a set of chemical reactions $\mathcal{R}$, and two species $\mathcal{S}_{j_I}$ (the input species) and $\mathcal{S}_{j_O}$ (the output species, with with $j_O\neq j_I$).
If the following three conditions hold:
\begin{enumerate}
    \item the R-graph of $\mathcal{R}$ admits a consistent labeling $\sigma$;
    \item the input species $\mathcal{S}_{j_I}$ is involved in exactly one reaction $\mathcal{R}_{i_I} \in \mathcal{R} $;
    \label{Icondition}
    \item the output species $\mathcal{S}_{j_O}$ is involved in exactly one reaction $\mathcal{R}_{i_O} \in \mathcal{R} $;\label{Ocondition}
\end{enumerate}
then, the species $\mathcal{S}_{j_O}$ is positively monotonic with respect to $\mathcal{S}_{j_I}$ if $\Gamma_{j_Ii_I}\sigma(i_I)$ and $\Gamma_{j_Oi_O}\sigma(i_O)$ have opposite signs, and negatively monotonic if they have the same sign. 
\end{Theorem}

We refer to~\cite{Nasti2020Robust} for the proof of Theorem~\ref{THEOREM}.


\subsection{The Input-Output monotonicity applied to the ERK signaling pathway} \label{example2}

Consider again the example of the ERK signaling pathway, described in Section~\ref{example}. 

We now apply the definition of the $\alpha$-robustness and we identify Raf and PPMek1 as the input and the output of the network. We assume that the initial  concentrations of all the species are fixed, but that the initial concentration of species Raf can vary from $1$ to $100$. 
 
In order to reduce the computational effort of simulation, we apply the sufficient condition of the Input-Output monotonicity. Indeed, if species PPMek1 is monotonic with respect to Raf, then just two simulations are necessary: one with Raf $=1$ and one with Raf $=100$. The dynamics of PPMek1 in all the other (intermediate) cases is included in the results we obtained from these two simulations.
 
\begin{figure}[t]
  \centerline{\includegraphics[width=0.9\linewidth]{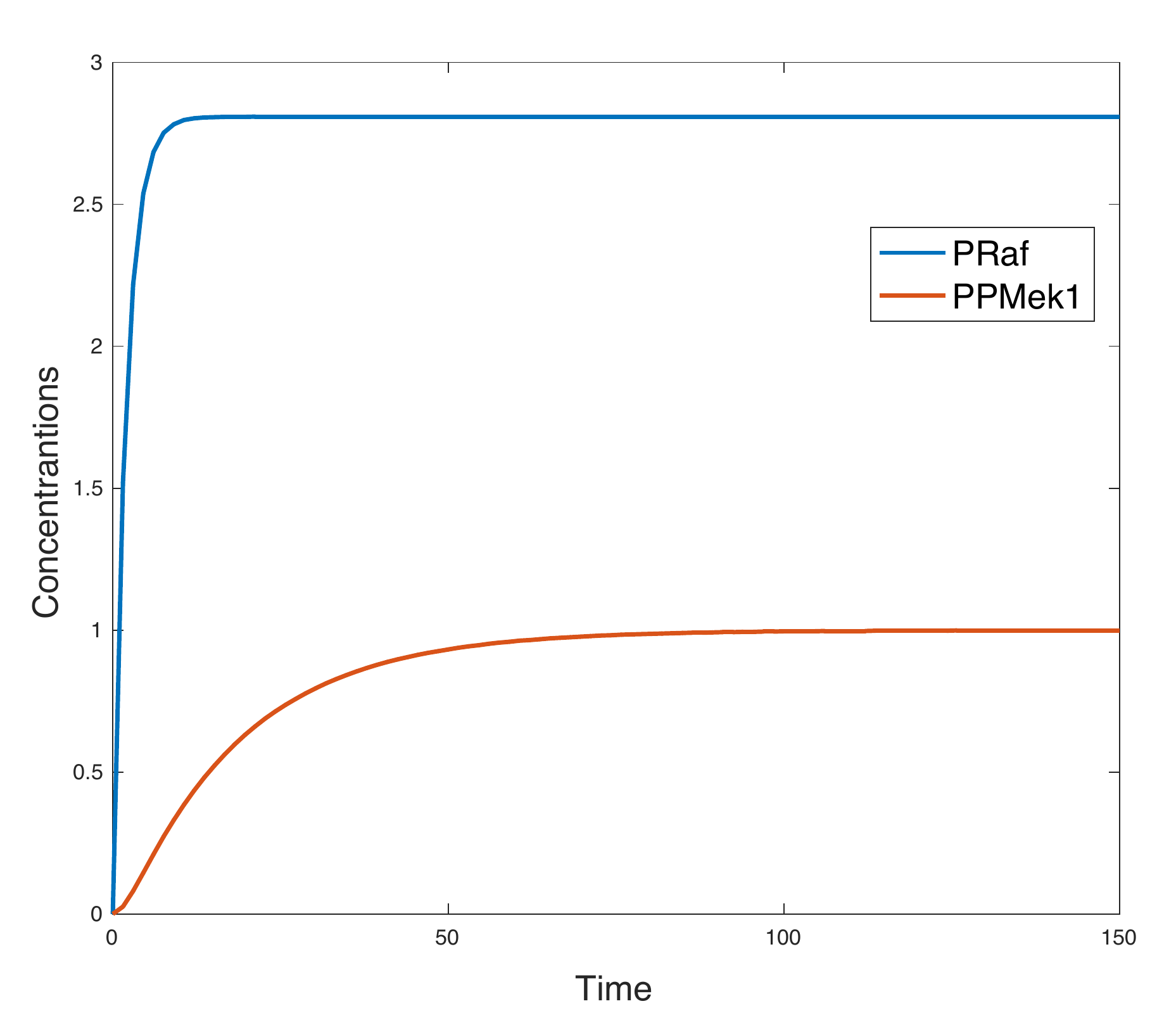}}
  \caption{
  Simulation results of Example \ref{erk_reaction1}, representing ERK signalling pathway. We show how the species PRaf reaches the steady state before than species PPMek1.}
      \label{Figure_4}
      \end{figure}

 As shown in detail in~\cite{nasti2020mono,Nasti2020Robust}, we notice that the species PRaf is the link between the two groups of reactions: in the first reaction, PRaf acts as the output of the network, while in the second one it acts as a kinetic constant. Since the reactions follow one another in a chain, we assume that \ch{PRaf} reaches the steady-state before the PPMek1 species, which represents the final product of the CRN, and we justify the assumption by simulations, as shown in Figure \ref{Figure_4}.

\begin{figure}[t]
 \centerline{\includegraphics[width=0.65\linewidth]{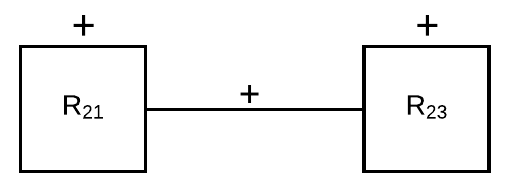}}
  \caption{Labeled R-graph of the second sub-network of CRN \ref{erk_reaction1}, representing ERK signalling pathway. Since the signs are both positive on the node-reactions, we can say that the output of the CRN, the species PPMek1, is positively monotonic w.r.t the input of the CRN, the species Mek1.}
      \label{Figure_5}
      \end{figure}

For the second sub-network it would be natural to select PRaf as the input of the network, since it represents the link between the first and the second block. However, we cannot choose PRaf as input, since it does not appear in the role of reactant or product. Therefore, we choose Mek1 as input. Indeed, we know that the concentration of $[PRaf]$ increases the rate of the reaction in which Mek1 is the only reactant. Consequently, to verify if PPMek1 is monotonic with respect to PRaf, we can verify if this species is monotonic with respect to the Mek1. Therefore, we choose Mek1 and PPMek1, respectively, as the input and the output. Then, we proceed by building the labelled R-graph, represented in Figure \ref{Figure_5}, and computing the following stoichiometric matrix:  

\[
\Gamma_{(Mek1,PPMek1),(\mathcal{R}_{21},\mathcal{R}_{23})} = \bordermatrix{
~ & \mathcal{R}_{21} & \mathcal{R}_{23} \cr
Mek1 & -1 & 0\cr
PPMek1 & 0 & +1\cr
}.
\]

By calculating the products $\Gamma_{j_Ii_I}\sigma(i_I)$ and $\Gamma_{j_Oi_O}s\sigma(i_O)$, which have opposite signs, we find that the species PPMek1 is positively monotonic with respect to the species Mek1. As consequence, since PRaf is positively monotonic with respect to Raf and PPMek1 is positively monotonic with respect to Mek1, PPMek1 is positively monotonic also with respect to \ch{Raf}, as we already noticed in Figure~\ref{Figure_6}.
  
This result enables us to use just two simulations (i.e., Raf$=1$ and Raf$=100$) to verify if the PPMek1 is robust with respect to the variation of the initial concentration of Raf.


\section{Conclusions} \label{conclusion}

Robustness is a crucial feature of many biological systems because this property allows their correct functioning in presence of molecular noise and environmental fluctuations. 

In order to verify the system's robustness, many strategies have been proposed. Among them, one of the most common approaches is to simulate the system with all possible combinations of initial concentrations of chemical species. 
However, this  requires a considerable number of simulations (in general an infinite number).  

In~\cite{rizk2009general}, Rizk et al. propose a general and computational framework for the definition of the robustness of biological functions with respect to a set of perturbations, based on LTL, an expressive language for specifying dynamical behaviors widely used in computer science and engineering.  Using this framework, implemented in BIOCHAM, they are able to describe on average how the system behaves under perturbations.  

In this paper, we show that we can cast our formal definition of the initial concentration robustness in the general framework proposed by Rizk et al.. Then, we can apply the sufficient condition of the Input-Output monotonicity in chemical reaction networks to substantially reduce the computational effort required by the huge number of simulations needed to prove the robustness property.   




%

\bibliographystyle{IEEEtran}
\bibliography{bib.bib}

\end{document}